\documentclass[sigconf]{acmart}
\usepackage{subfigure} 
\usepackage[ruled]{algorithm2e}
\usepackage{algorithmic}
\usepackage{threeparttable}
\usepackage{wasysym}
\usepackage{enumitem}
\usepackage{multirow}
\usepackage{balance}
\usepackage[table]{xcolor}
\AtBeginDocument{%
  }

\copyrightyear{2026}
\acmYear{2026}
\setcopyright{cc}
\setcctype{by}
\acmConference[KDD '26]{Proceedings of the 32nd ACM SIGKDD Conference on Knowledge Discovery and Data Mining V.1}{August 09--13, 2026}{Jeju Island, Republic of Korea}
\acmBooktitle{Proceedings of the 32nd ACM SIGKDD Conference on Knowledge Discovery and Data Mining V.1 (KDD '26), August 09--13, 2026, Jeju Island, Republic of Korea}
\acmPrice{}
\acmDOI{10.1145/3770854.3780284}
\acmISBN{979-8-4007-2258-5/2026/08}

\begin{document}

\title{ProEx: A Unified Framework Leveraging Large Language Model with Profile Extrapolation for Recommendation}

\author{Yi Zhang}
\authornote{The work was done while the author was visiting the University of Queensland.}
\affiliation{%
  \institution{Anhui University}
  \city{Hefei}
  \country{China}}
\email{zhangyi.ahu@gmail.com}

\author{Yiwen Zhang}
\authornote{Yiwen Zhang and Hongzhi Yin are co-corresponding authors.}
\affiliation{%
  \institution{Anhui University}
  \city{Hefei}
  \country{China}
}
\email{zhangyiwen@ahu.edu.cn}

\author{Yu Wang}
\affiliation{%
  \institution{Anhui University}
  \city{Hefei}
  \country{China}
}
\email{wangyuahu@stu.ahu.edu.cn}

\author{Tong Chen}
\affiliation{%
  \institution{The University of Queensland}
  \city{Brisbane}
  \country{Australia}
}
\email{tong.chen@uq.edu.au}

\author{Hongzhi Yin}
\authornotemark[2]
\affiliation{%
  \institution{The University of Queensland}
  \city{Brisbane}
  \country{Australia}
}
\email{h.yin1@uq.edu.au}

\renewcommand{\shortauthors}{ Yi Zhang, Yiwen Zhang, Yu Wang, Tong Chen, \& Hongzhi Yin}

\begin{abstract}
The powerful text understanding and generation capabilities of large language models (LLMs) have brought new vitality to general recommendation with implicit feedback. One possible strategy involves generating a unique user (or item) profile from historical interaction data, which is then mapped to a semantic representation in the language space.  However, a single-instance profile may be insufficient to comprehensively capture the complex intentions behind a user's interacted items. Moreover, due to the inherent instability of LLMs, a biased or misinterpreted profile could even undermine the original recommendation performance. Consequently, an intuitive solution is to generate multiple profiles for each user (or item), each reflecting a distinct aspect of their characteristics. In light of this, we propose a unified recommendation framework with multi-faceted profile extrapolation (\textsf{ProEx}) in this paper. By leveraging chain-of-thought reasoning, we construct multiple distinct profiles for each user and item. These new profiles are subsequently mapped into semantic vectors, extrapolating from the position of the original profile to explore a broader region of the language space. Subsequently, we introduce the concept of environments, where each environment represents a possible linear combination of all profiles. The differences across environments are minimized to reveal the inherent invariance of user preferences.
We apply \textsf{ProEx} to three discriminative methods and three generative methods, and conduct extensive experiments on three datasets. The experimental results demonstrate that \textsf{ProEx} significantly enhances the performance of these base recommendation models.
\end{abstract}




\maketitle

\section{Introduction}
General recommender systems \cite{ricci2011introduction} infer implicit user preferences from historical interactions \cite{rendle2009bpr}, primarily employing collaborative filtering \cite{sarwar2001item, he2017neural} based solely on ID-level co-occurrence without integrating semantic content \cite{yuan2023go}. At present, 
the recommendation task increasingly revolves around learning fixed-dimensional representations for users and items in a unified latent space \cite{rendle2009bpr, yin2025device}, encompassing discriminative models \cite{he2020lightgcn, yu2022graph} that encode semantic embeddings and generative methods \cite{liang2018variational, wang2017irgan} that model preference distributions probabilistically. However, the sparsity of interactions limits the supervision available for learning \cite{yin2020overcoming}, hindering the recommender's ability to extract reliable collaborative signals \cite{wang2019neural}.

\begin{figure}[!t]
\setlength{\abovecaptionskip}{0.1cm}
\setlength{\belowcaptionskip}{0.1cm} 
  \centering
  \includegraphics[width=\linewidth]{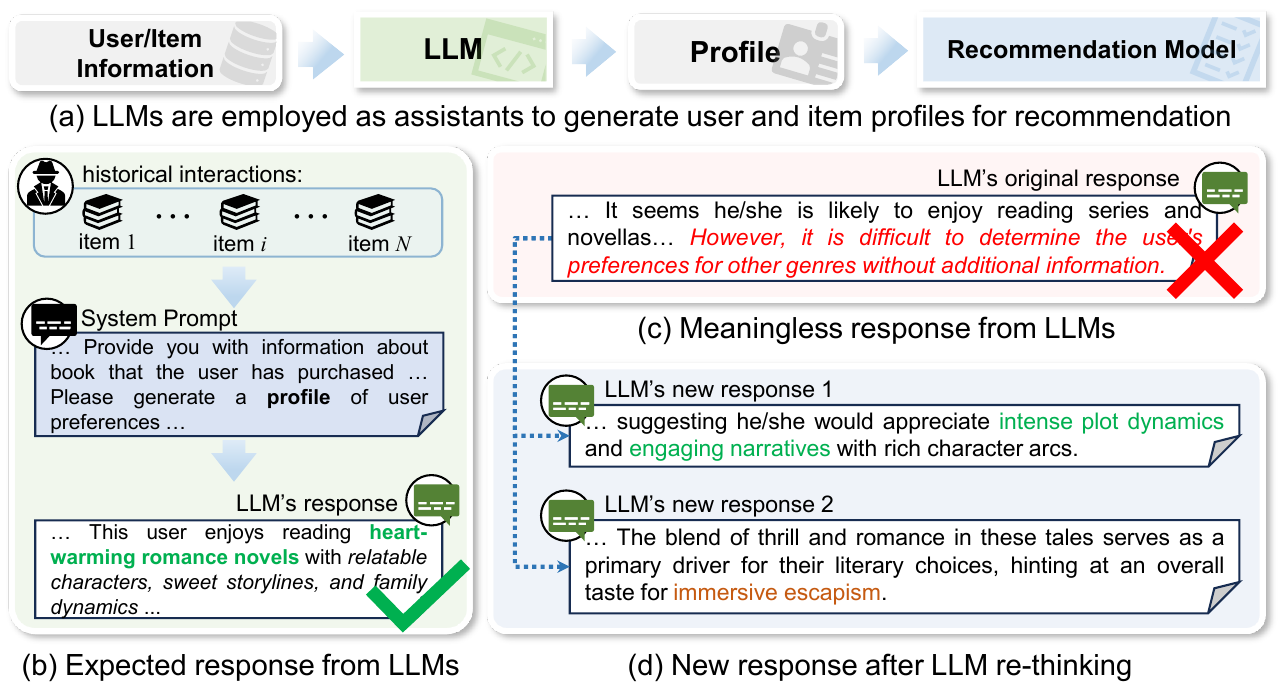}
  \caption{The process of using LLM as a recommendation assistant for profile generation. A  single-instance profile for each user and item may contain errors and biases.}
\label{fig_motivation}
\end{figure}

Recently, pre-trained large language models (LLMs) \cite{touvron2023llama, chang2024survey}, empowered by extensive textual corpora and advanced generative capacities \cite{zhao2023survey}, exhibit remarkable generalization ability. Capitalizing on this capability, researchers have explored leveraging LLMs in recommender systems as a potential remedy for data sparsity.
The first line of work involves fine-tuning LLMs \cite{hulora2022, zhao2023survey} to act as a holistic recommender \cite{geng2022recommendation, liao2024llara}, which however incurs high computational costs \cite{zhang2023recommendation} and tends to suffer from poor generalizability across varying recommendation scenarios \cite{liao2024llara}. An alternative strategy is to leverage LLMs as a plug-and-play enhancement \cite{xi2024towards, zhang2024agentcf} for base recommendation models by incorporating rich semantic priors \cite{yin2014temporal, wang2025lettingo}. As depicted in Fig. \ref{fig_motivation}(a), to harness their advantages in text comprehension and generation, LLMs are employed to encode interactions and side information into compact representations (\textit{e.g.}, profiles \cite{zhang2024generative, ren2024representation}). Thus, this paradigm fully leverages salient behavioral or semantic features \cite{wang2025unleashing, zhang2025towards}, achieving encouraging empirical results while being more flexible and efficient \cite{yin2015dynamic}.

However, LLM-based profiling still faces unresolved limitations. On the one hand, representing each user or item with a single profile often overlooks the complex and multi-faceted nature of user behaviors and item attributes \cite{wang2021learning, zhang2024exploring}. On the other hand, the reliability of LLM-generated outputs remains a significant concern: LLMs are susceptible to issues such as noise \cite{wei2024llmrec}, hallucination \cite{ji2023survey}, popularity bias \cite{zhang2023chatgpt},  and semantic inconsistency \cite{radford2021learning}, which may result in inaccurate, inconsistent, or semantically irrelevant profiles. As shown in Fig. \ref{fig_motivation}(b), our objective is to use LLMs to build a high-quality profile that reflects user preferences for downstream recommendation. Consequently, when user-item interactions exhibit high sparsity or complex patterns, one-shot profiling may yield unreliable or meaningless responses, as illustrated in Fig. \ref{fig_motivation}(c). Such profiles may misrepresent user interests and introduce noise during training \cite{wang2025unleashing}, impeding the subsequent recommendation performance. The unreliability of a single profile, coupled with the difficulty of directly evaluating the quality of each generated profile, makes the situation even more difficult. Under these conditions, any imperfections present in LLM-generated profiles could be inherited and magnified by downstream recommendation models.


This observation provides a natural motivation for the idea: \textit{instead of depending on a single profile, \textbf{can we generate multiple diverse profiles per user or item} to more effectively capture their complex and multi-faceted characteristics from complementary perspectives? } Intuitively, leveraging the collective intelligence of multiple diverse profiles not only enhances downstream recommendation performance by enriching the semantic representations of users and items, but also provides a potential remedy to noisy or biased profiles as the recommendation process is now conditioned on a combination of diverse profiles instead of a single-minded one. Nevertheless, this paradigm brings forth two fundamental challenges that require thorough investigation and resolution:
\begin{itemize}[leftmargin=*]
\item[$\bullet$] The first challenge involves generating multiple semantically diverse profiles that encapsulate complementary facets of user behaviors or item characteristics.

\item[$\bullet$] The second challenge lies in integrating diverse profiles into the base recommender while preserving their maximum relevance to user preferences, especially when profile quality cannot be directly assessed.
\end{itemize}

To tackle the above challenges, we propose a novel unified LLM-enhanced Framework with Profile Extrapolation for general recommendation (\textbf{\textsf{ProEx}}). Specifically, given a user's (or item's) interaction history and side information (\textit{e.g.}, tags, descriptions, or reviews), we initially employ an LLM to derive a original profile. A chain-of-thought reasoning procedure is subsequently introduced to extract key information and iteratively re-generate multi-faceted profiles under a low-similarity constraint, as shown in Fig. \ref{fig_motivation}(d). This ensures the profiles generated about each user/item are grounded and diverse. From a feature space perspective, our goal is to improve the uniformity of the language-induced semantic space, enabling user (item) preferences (characteristics) to generalize toward distant regions. We term this process multi-faceted \textbf{profile extrapolation}. Afterwards, these profiles are projected into embedding representations and fused with the base recommender through cross-space alignment. To leverage the multi-faceted profiles effectively while reducing bias, we introduce the concept of environments, where profiles are mixed at varying ratios within each environment. Then, by constraining discrepancies across environments, invariant profile features underlining a user or item's key characteristics are extracted. Finally, we further amplify the profile extrapolation through a contrastive regularization process, enabling user preferences to cover a broader range of possibilities. The proposed \textsf{ProEx} framework is unified and compatible with both discriminative and generative recommendation models. The major contributions of this paper are summarized as follows:

\begin{itemize}[leftmargin=*]
\item[$\bullet$] We propose a unified recommendation framework called \textsf{ProEx}, which provides multiple multi-faceted profiles for users (or items) through chain-of-thought reasoning, thus mitigating the limitations of existing LLM-based profiling for recommendation.
\item[$\bullet$] We introduce the concept of environments for multiple profiles, where profiles are mixed at varying ratios in each environment to mitigate bias, and invariant features are extracted by minimizing intra-environment discrepancies.
\item[$\bullet$] We apply \textsf{ProEx} to three discriminative methods and three classic generative methods on three public datasets. The results indicate that \textsf{ProEx} substantially improves base model performance and outperforms existing LLM-based recommendation approaches.
\end{itemize}

\section{Methodology}
\subsection{Problem Formulation}
\begin{figure*}[t]
\setlength{\abovecaptionskip}{0.1cm}
\setlength{\belowcaptionskip}{0.1cm} 
  \centering
  \includegraphics[width=\linewidth]{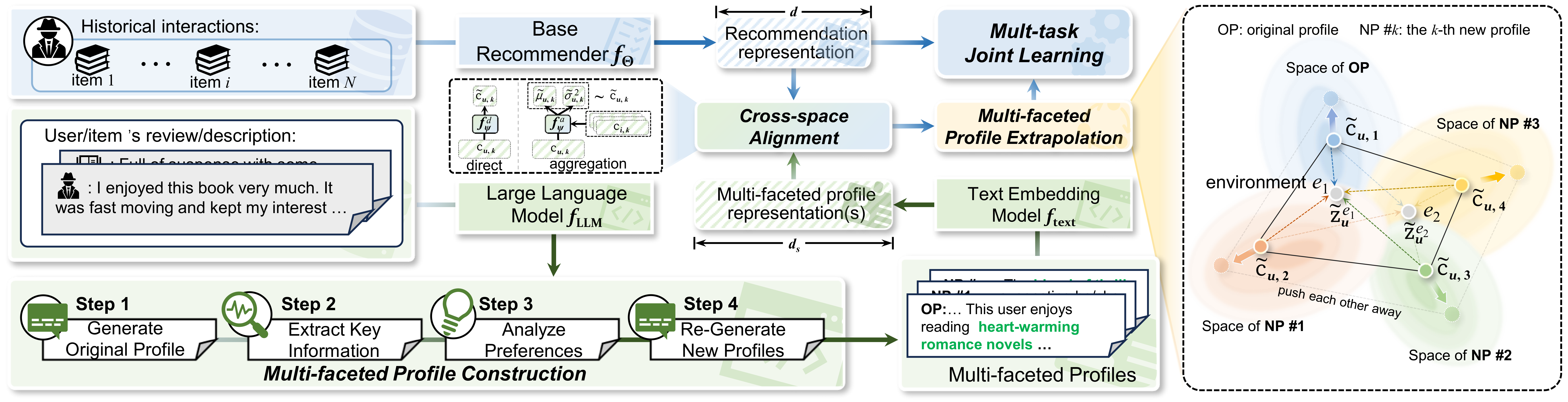}
  \caption{The information flow of the proposed \textsf{ProEx}.}
\label{fig_model}
\end{figure*}

In a general ID-based recommendation scenario, we consider $M$ users ($\mathcal U=\{u_1, u_2, ..., u_M\}$), $N$ items ($\mathcal I=\{i_1, i_2, ..., i_N\}$), and the interactions between them \cite{liang2018variational, he2020lightgcn}. For any user $u \in \mathcal U$, the historical interactions are represented by an interaction vector $\mathbf x_u \in \mathbb R^{1 \times N}$. If user $u$ has interacted with item $i$, the corresponding value in the vector $x_{ui}$ is set to 1. The goal of the recommender system is to learn a prediction model $f_\Theta$ that estimates the preference score $\hat{x}_{uj}$ for all unobserved items $\{j \in \mathcal I / i\}$ that the user has not interacted with. Based on this, we propose the model-agnostic recommendation framework \textsf{ProEx}, as illustrated in Fig. \ref{fig_model}.

\subsection{Base Recommender}
 Depending on the modeling strategy, ID-based recommender systems are generally divided into two categories: discriminative and generative methods \cite{yu2023self, wang2023diffusion}.

\subsubsection{\textbf{Discriminative Method}} Discriminative methods \cite{rendle2009bpr, he2017neural} typically map user and item IDs to a shared collaborative space $\mathcal X \in \mathbb R^{d}$, with the core idea being to learn decision boundaries that distinguish different user preferences, allowing for predictions on unobserved items \cite{he2017neural, he2020lightgcn}. Specifically, given user $u$ and item $i$, their corresponding embedding representations are denoted as $\mathbf z_u \in \mathbb R^{1\times d}$ and $\mathbf z_i \in \mathbb R^{1\times d}$, respectively. Formally, representations $\mathbf z_u$ and $\mathbf z_i$ are initially drawn from a predefined prior $p(\mathbf z_u,\mathbf z_i)$. The joint posterior distribution over these latent representations is then inferred conditioned on the observed user–item interactions $\mathbf x_u$:

\begin{equation}
p(\mathbf z_u,\mathbf z_i|\mathbf x_u)\propto p(\mathbf x_u|\mathbf z_u,\mathbf z_i)\cdot p(\mathbf z_u,\mathbf z_i),
\end{equation}
where $p(\mathbf x_u|\mathbf z_u,\mathbf z_i)$ denotes the likelihood of generating the observed data $\mathbf x_u$, which serves as the objective to be maximized during the training process \cite{rendle2009bpr}. The user’s preference for the item is then estimated based on the similarity between their embeddings.

\subsubsection{\textbf{Generative Method}} Given any user $u$, generative methods \cite{yu2019vaegan, wang2023diffusion} commonly  operate under the assumption that the observed data point $\mathbf x_u$ is sampled from an underlying joint distribution $p(\mathbf x_u, \mathbf z_u)$, where $\mathbf z_u \in \mathbb R^{1\times d}$ is a latent continuous variable in the collaborative space $\mathcal X \in \mathbb R^{d}$. Generative modeling aims to maximize likelihood $p(\mathbf x_u)$ by directly marginalizing the latent variable $p(\mathbf x_u) =\int p(\mathbf x_u, \mathbf z_u)d\mathbf z_u $ \cite{luo2022understanding}. Given the intractability of directly solving the integral, taking the variational auto-encoder (VAE) \cite{liang2018variational, zhang2023revisiting} as an example, an approximate posterior distribution $q_{\phi}(\mathbf z_u|\mathbf x_u)$ is typically introduced:
\begin{equation}
\label{vae_q}
q_{\phi}(\mathbf z_u|\mathbf x_u)=\mathcal N\left (\mathbf z_u|\boldsymbol\mu_u, \text{diag}[\boldsymbol\sigma_u^2]\right),
\end{equation}
where $\boldsymbol\mu_u$ and $\boldsymbol\sigma_u$ are the non-linear mean and variance of user $u$ parameterized by $\phi$, respectively. By employing the reparameterization trick \cite{kingma2013auto}, an approximate variational distribution $q_{\phi}(\mathbf z_u|\mathbf x_u)$ over the latent $\mathbf z_u$ is constructed, which characterizes the user's preference over the entire item set $\mathcal I$ \cite{liang2018variational}, as opposed to learning a fixed embedding for each user as in discriminative methods \cite{he2020lightgcn}.

Regardless of the modeling paradigm, the objective remains to learn a semantic representation $\mathbf z_u$ for user $u$\footnote{To maintain consistency with the generative methods,  in the following we focus on describing user representation $\mathbf z_u$ in discriminative methods and omit the item side $\mathbf z_i$, as the modeling processes are analogous.}. Therefore, the recommendation process can be carried out either by directly computing the similarity between user and item representations $\mathbf z_u$ and $\mathbf z_i$ \cite{rendle2009bpr, zhang2020gcn}, or by employing a decoder $f_\theta$ parameterized by $\theta$ to reconstruct user–item interactions based on latent variable $\mathbf z_u$ \cite{liang2018variational}:
\begin{equation}
\label{score}
p(x_{u,i}|\mathbf z_u, \mathbf z_i) = \delta(\mathbf z_u^\top \mathbf z_i); \quad p_\theta(\mathbf x_u|\mathbf z_u) \sim \delta(f_\theta(u)),
\end{equation}
where $\delta$ is an activation function, which regulates the results in the range of 0 to 1.

\begin{figure}[t]
\setlength{\abovecaptionskip}{0.1cm}
\setlength{\belowcaptionskip}{0.1cm} 
  \centering
  \includegraphics[width=\linewidth]{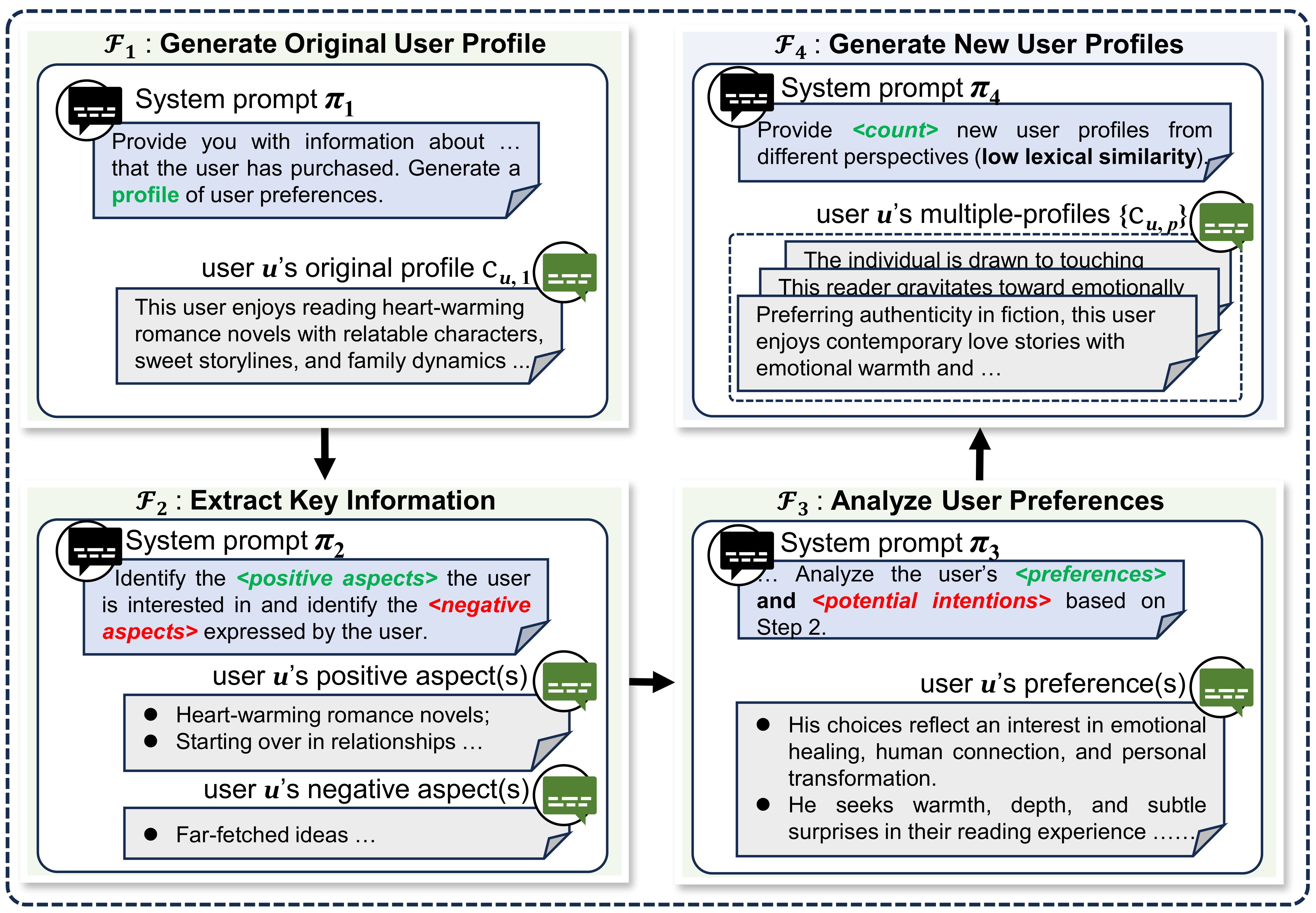}
  \caption{The multi-faceted profile construction process for user side based on four-step reasoning (due to the space limitations, only key prompt words are shown in the figure).}
    \label{fig_cot}
\end{figure}

\subsection{Multi-faceted Profile Extrapolation}
Given any large language model (LLM) denoted by $f_{\text{LLM}}$, profile generation can be facilitated by system prompts to summarize historical interactions, user reviews, and item attributes into fixed-length text description \cite{ren2024representation, zhao2025synthetic}. In this section, we aim to construct a set of multi-faceted profiles for each user $u$ (and item $i$) that can capture a diversity of aspects.

\subsubsection{\textbf{Multi-faceted Profile Construction}}
Given user $u$'s interactions $\mathbf x_u$ and auxiliary information $\mathbf r_u$ (\textit{e.g.}, titles and reviews of interacted items), we define a four-stage chain-of-thought \cite{wei2022chain} reasoning procedure composed of a sequence of transformation functions $\{\mathcal F_1, \mathcal F_2, \mathcal F_3, \mathcal F_4\}$, guided by a set of corresponding system prompts $\{\pi_1, \pi_2, \pi_3, \pi_4\}$:
\begin{equation}
\label{CoT}
  \mathcal P_u = \mathcal F_4^{\pi_4} \circ \mathcal F_3^{\pi_3} \circ \mathcal F_2^{\pi_2} \circ \mathcal F_1^{\pi_1}(\mathbf x_u,\mathbf r_u).
\end{equation}
The corresponding construction process is illustrated in Fig. \ref{fig_cot}. In $\mathcal F_1$, we input the textual information associated with the user $u$ to obtain the \textbf{original profile (OP)}. In $\mathcal F_2$, information from the OP is extracted to identify the user $u$’s positive and negative aspects toward interacted items. Subsequently, in $\mathcal F_3$, the user $u$’s latent preferences are further analyzed. All the above information is then leveraged in $\mathcal F_4$ to generate multiple \textbf{new profiles (NPs)}. 

It is important to note that, since we aim to characterize the user from diverse perspectives and mitigate the representational collapse caused by popularity bias, we require the LLM in $\mathcal F_4$ to generate NPs that differ entirely from OP in terms of wording and expression. Upon completing the reasoning process, we can obtain the user $u$’s profile set $\mathcal P_u=\{s_{u,1}, s_{u,2}, ..., s_{u,K}\}$, where $K=|\mathcal P_u|=|\mathcal P_i|$ is the number of profiles. $s_{u,1}$ denotes user $u$'s OP,  and $\{s_{u,2}, ..., s_{u,K}\} $ correspond to the $K-1$ NPs of user $u$. For the item side, we have a similar definition for $\mathcal P_i=\{s_{i,1}, s_{i,2}, ..., s_{i,K}\}$.

After constructing multiple user profiles, it is necessary to further transform these textual descriptions into semantically rich representations, enabling collaborative learning with the corresponding representation $\mathbf z_u$ (or $\mathbf z_i$) in the collaborative space $\mathcal X$ \cite{ren2024representation}. Therefore, a text embedding model $f_{\text{text}}$ \cite{su2023one} is employed to transform all user $u$'s profiles into a fixed-dimension semantic representation:
\begin{equation}
\label{trans}
\mathbf C_u = \{\mathbf c_{u,1}, \mathbf c_{u,2}, ..., \mathbf c_{u, K}\}, \ \text{where} \ \mathbf c_{u,k} = f_{\text{text}}(s_{u,k}) \in \mathbb R^{1\times d_s}. 
\end{equation}
It is worth noting that the semantic representation $\mathbf c_{u,k}$ from $f_{\text{text}}$ typically has much higher dimensionality than those used in the base recommendation model $f_\Theta$ (\textit{i.e.,} $d_s \gg d$) \cite{ren2024representation}, and resides in a distinct language space $\mathcal Y \in \mathbb R^{d_s}$ \cite{zhang2025towards}. Therefore, we introduce a mapping function $f_\psi : \mathcal X \times \mathcal Y \to \mathcal X$ parameterized by $\psi$ to perform cross-space dimension alignment. Considering that generative models typically do not explicitly model item-side representations \cite{liang2018variational, wang2023diffusion}, it is necessary to handle this case separately. Building on this, we adopt direct alignment $f_\psi^d$ for discriminative methods and aggregation alignment $f_\psi^a$ for generative methods:

\begin{equation}
\begin{aligned}
\label{align}
\text{direct}: &\tilde{\mathbf c}_{u,k}= f_{\psi}^d(\mathbf c_{u,k}), \quad \tilde{\mathbf c}_{i,k}= f_{\psi}^d(\mathbf c_{i,k});\\
\text{aggregation}: &\tilde{\mathbf c}_{u,k} \sim f_{\psi}^a (\text{agg}( \{\mathbf c_{i,k}|, i \in \mathcal M(u)\}) + \mathbf c_{u,k}),
\end{aligned}
\end{equation}
where the aggregation function $\text{agg}(\cdot)$ integrates information from all items in the neighbor set $\mathcal M(u)$ of user $u$. For discriminative methods, we set $f_\psi^d: \mathbb{R}^{d_s} \to \mathbb{R}^{d}$ and can get the profile representations $\tilde{\mathbf c}_{u,k}$ and $\tilde{\mathbf c}_{i,k}$ directly; for generative methods, we have $f_\psi^a: \mathbb{R}^{d_s} \to \mathbb{R}^{2d}$, and the output of the aggregation mapping function $f_{\psi}^a$ is evenly split into two parts, which are used as the mean $\boldsymbol\mu_{u,k} \in \mathbb R^{d}$ and variance $\boldsymbol\sigma^2_{u,k} \in \mathbb R^{d}$, thereby constructing the approximate posterior $\tilde{\mathbf c}_{u,k} \sim\mathcal N(\boldsymbol\mu_{u,k},  \boldsymbol\sigma^2_{u,k}) \in \mathbb R^{d}$ for the $k$-th profile of user $u$ with sampling \cite{kingma2013auto, yin2019social}. 

\begin{figure}[t]
\setlength{\abovecaptionskip}{0.1cm}
\setlength{\belowcaptionskip}{0.1cm} 
  \centering
  \includegraphics[width=\linewidth]{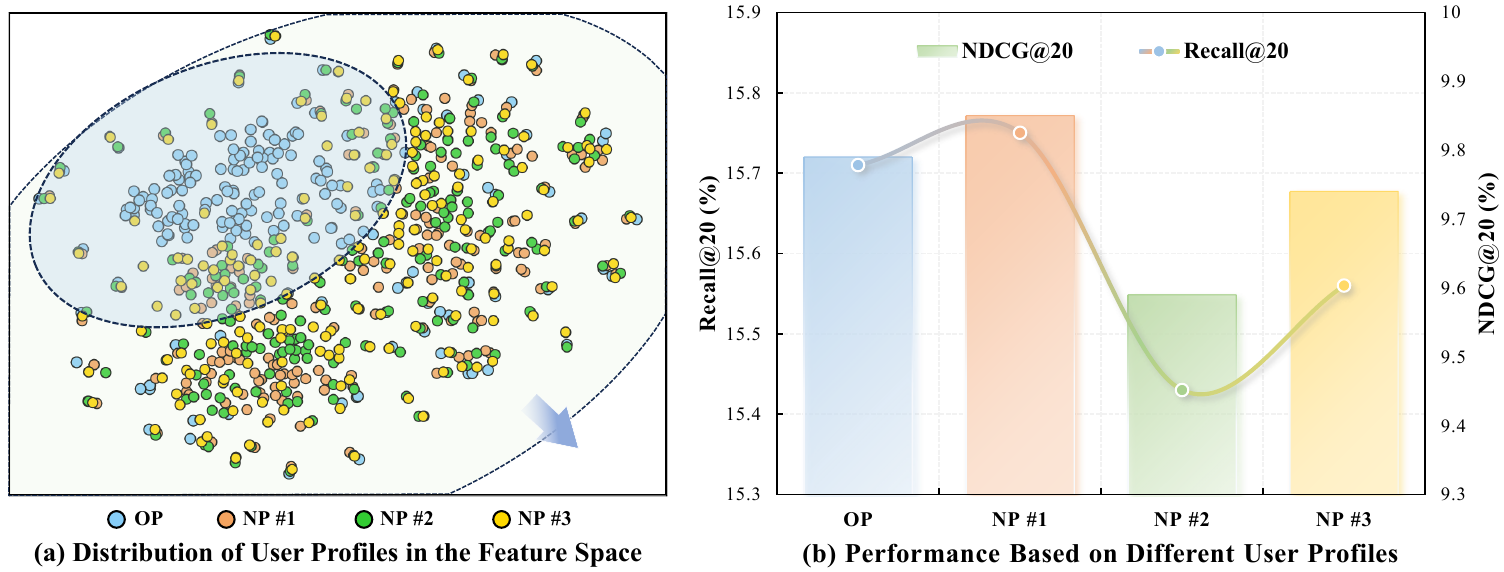}
  \caption{(a) The distribution of the original profile (OP) and the three new profiles (\textbf{NP \#1, NP \#2}, and \textbf{NP \#3}) in the feature space; (b) The performance of different profiles on the base model Mult-VAE \cite{liang2018variational} on Amazon-Book dataset \cite{ren2024representation}.}
\label{fig_new_profiles}
\end{figure}

\subsubsection{\textbf{Profile Extrapolation}} 
In practice, we perform a quantitative analysis with $K=4$, where an original profile (\textit{i.e.}, OP) and three newly generated profiles (denoted as \textbf{NP \#1, NP \#2}, and \textbf{NP \#3}) are obtained for each user based on multi-aspect profile construction. Fig. \ref{fig_new_profiles}(a) illustrates the distribution of the original and newly generated profiles in the language space $\mathcal Y$. We observe that the original profiles (blue dots) are predominantly clustered around a central region (blue circle), reflecting a strong popularity bias. In contrast, the newly generated profiles are more widely dispersed throughout the feature space, minimizing their similarity to the original profiles. This observation demonstrates the effectiveness of the proposed CoT-based profile construction process. By occupying a broader region and expanding the feature space boundary, this process gives rise to \textbf{profile extrapolation}.

Considering that our goal is to ensure diversity among different profiles in order to characterize the user $u$ from multiple perspectives, we further introduce the contrastive regularization \cite{zhang2024exploring} to push apart these profiles in the feature space:
\begin{equation}
\label{reg}
\mathcal L_{\text{reg}}=\sum_{k=1}^{K}\text{log}\left (1+\text{exp}(1/\tau) \sum_{k'=1,k'\ne k}^{K}\text{exp}\left (\tilde{\mathbf c}_{u,k}^\top \tilde{\mathbf c}_{u,k'}/\tau \right )\right ),
\end{equation}
where $\tau$ is a predefined temperature coefficient used to adjust the uniformity \cite{wang2020understanding} of the feature space (set to 0.2 by default \cite{yu2022graph}). Optimization gradually enlarges the pairwise distances between profiles. As depicted by the dashed background in the right panel of Fig. \ref{fig_model}, this expansion of the convex hull facilitates further profile extrapolation of user preferences.

Given the difficulty of evaluating the quality of profiles, it remains uncertain whether all newly generated profiles positively contribute to the recommendation. Based on this concern, we adopt Mult-VAE \cite{liang2018variational} as the base model and incorporate each profile individually into its training process on Amazon-Book dataset \cite{zhang2025towards}, as shown in Fig. \ref{fig_new_profiles}(b). As is evident, there exists a substantial performance disparity among different profiles, with NP \#2 even leading to a significant degradation in recommendation performance. Combined with Fig. \ref{fig_new_profiles}(a), although we explicitly enforce reduced semantic similarity in the newly generated profiles, this process may inadvertently compromise the original semantic information. Therefore, it is necessary to reconsider how to effectively integrate these profiles $\{\tilde{\mathbf c}_{u,k}\}_{k=1}^{L}$  for recommendation task.

Although these profiles differ from one another, they essentially describe the same user $u$ and therefore should remain consistent with respect to the user’s preferences. Inspired by the theory of invariant learning \cite{bai2024multimodality, krueger2021out}, we introduce the concept of \textbf{environments} to collaboratively leverage these new profiles. Given any environment $e \in \mathcal E$, for discriminative methods,  we have:
\begin{equation}
\label{mix_1}
\tilde{\mathbf c}_{u}^e=\vartheta_{1}^e\tilde{\mathbf c}_{u,1} +  \vartheta^e_{2}\tilde{\mathbf c}_{u,2} + ...+\vartheta^e_{K}\tilde{\mathbf c}_{u,K}; \quad\text{s.t.} \sum_{k =1}\vartheta_k^e=1, 
\end{equation}
where LLM-generated profiles $\{\tilde{\mathbf c}_{u,k}\}_{k=1}^{K}$ are weighted by $\{\vartheta^e_{k}\}_{k=1}^{K}$. 

For generative methods, according to Eq. \ref{align}, we model the mean $\tilde{\boldsymbol\mu}_{u,k}$ and variance $\tilde{\boldsymbol\sigma}^2_{u,k}$ over each user $u$'s $K$ profiles. Therefore, we consider them separately:
\begin{equation}
\label{mix_2}
\tilde{\boldsymbol\mu}_{u}^e=\vartheta_{1}^e\tilde{\boldsymbol\mu}_{u,1} +  \vartheta^e_{2}\tilde{\boldsymbol\mu}_{u,2} + ...+\vartheta^e_{K}\tilde{\boldsymbol\mu}_{u,K}; \quad\text{s.t.} \sum_{k=1}\vartheta_k^e=1,
\end{equation}
where the variance $(\tilde{\boldsymbol\sigma}^e_{u,k})^2$ uses a similar definition, such that we can also obtain the result $\tilde{\mathbf c}_{u}^e \sim \mathcal N(\tilde{\boldsymbol\mu}_{u}^e, (\tilde{\boldsymbol\sigma}_{u}^{e})^2)$. 

In each environment $e$, the construction process for discriminative and generative methods is identical. Concretely, given a fixed set of profiles, their respective contributions are controlled by a set of coefficients $\{\vartheta^e_{k}\}_{k=1}^{K}$:
\begin{equation}
\label{Dirichlet}
\{\vartheta_{1}^e, \vartheta_{2}^e, ...,\vartheta_{K}^e\} \sim \text{Dirichlet}(\alpha_1,\alpha_2, ..., \alpha_{K}),
\end{equation}
where $\{\alpha_k\}$ is a set of shape parameters \cite{zhang2025mixrec} in the Dirichlet distribution. Here, we introduce a sampling-based strategy to generate entirely different combinations of coefficients $\{\vartheta^e_{k}\}_{k=1}^{K}$ in each environment $e$, thereby introducing variability across environments to enhance generalization. 

As illustrated on the far right of Fig. \ref{fig_model}, we set up two environments ($e_1$ and $e_2$) as an example. By combining the coefficients $\{\vartheta^e_k\}$ differently, we can easily obtain two new representations $\tilde{\mathbf c}_{u}^{e_1}$ and $\tilde{\mathbf c}_{u}^{e_2}$ in linear time complexity, positioned in distinctly different locations. This convex combination ensures that the newly generated profiles $\tilde{\mathbf c}_{u}^{e_1}$ and $\tilde{\mathbf c}_{u}^{e_2}$ always lie within the convex hull formed by the original profile $\tilde{\mathbf c}_{u,1}$ and three new profiles $\{\tilde{\mathbf c}_{u,2}, \tilde{\mathbf c}_{u,3}, \tilde{\mathbf c}_{u,4}\}$. Based on this, we not only achieve an adaptive fusion of multiple profiles, but also fully leverage the semantic space spanned by them. 

{\small
\begin{algorithm}
\caption{The training process of \textsf{ProEx}}
\label{algorithms}
\KwIn{user–item ineraction $\mathbf X$, large language model $f_{\text{LLM}}$, text embedding model $f_{\text{text}}$, base recommendation model $f_{\Theta}$, mapping function $f_\psi$, fusion function $f_\gamma$, transformation functions $\{\mathcal F_1, \mathcal F_2, \mathcal F_3, \mathcal F_4\}$, and system prompts $\{\pi_1, \pi_2, \pi_3, \pi_4 \}$.}
\begin{algorithmic}[1]
\STATE initialize model parameters for $f_{\Theta}$ and $f_\psi$;
\STATE retrieve all multi-faceted user/item profiles $\mathcal P_u$ ($\mathcal P_i$) by $f_{\text{LLM}}$ with four-step chain-of-thought reasoning  (Eq . \ref{CoT});
\WHILE {\textsf{ProEx} not converge}
\STATE sample a mini-batch of user-item pairs $\mathcal B$;
\FOR{$u\in \mathcal B$}
\STATE construct recommendation representation $\mathbf z_u$ by $f_{\Theta}$;
\FOR{$k= 1, 2, ... \mathcal K$}
\STATE  transform the profile $s_{u,k}$ into semantic representation $\mathbf c_{u,k}$ by $f_{\text{text}}$ (Eq. \ref{trans});
\STATE construct dimension-aligned representation $\tilde{\mathbf c}_{u,k}$ by $f_{\psi}$ (Eq. \ref{align});
\ENDFOR
\STATE calculate the contrastive regularization loss $\mathcal L_{\text{reg}}$ by Eq. \ref{reg};
\FOR{$e= 1, 2, ... \mathcal E$}
\STATE  sample $\{\vartheta_{1}^e, \vartheta_{2}^e, ...,\vartheta_{K}^e\}$ from $\text{Dirichlet}(\alpha_1,\alpha_2, ..., \alpha_{K})$ (Eq. \ref{Dirichlet});
\STATE  construct mixed profile representation $\tilde{\mathbf c}_{u}^e$ by Eq. \ref{mix_1} or Eq. \ref{mix_2};
\STATE  construct fused representation ${\mathbf z}_{u}^e$ by Eq. \ref{fuse};
\STATE  calculate the environment-specific loss $\mathcal L^e$ by Eq. \ref{env_loss};
\ENDFOR
\STATE calculate the total loss and variance over all environments by Eq. \ref{loss};
\ENDFOR
\STATE average gradients from mini-batch;
\STATE update model parameters by descending the gradients $\nabla_{\Theta, \psi}\mathcal L_{\textsf{ProEx}}$;

\ENDWHILE
\RETURN model parameters $\Theta, \psi$;
\end{algorithmic}
\end{algorithm}
}

\subsection{\textsf{ProEx} Framework}
For user $u$, we have obtained representations $\mathbf z_u$ and $\{\tilde{\mathbf c}_{u}^e|e\in \mathcal E\}$ with identical dimensionality, derived respectively from the collaborative space $\mathcal X$ (via the recommendation model $f_{\Theta}$) and the language space $\mathcal Y$ (via the LLM $f_{\text{LLM}}$). Due to semantic discrepancies between the spaces \cite{zhang2025towards}, as well as differences in probability density and distributional shape \cite{xu2020learning}, directly fusing these representations can result in undesirable information loss \cite{ren2024representation}. Therefore, it is essential to perform semantic alignment between the two spaces. 

Since we do not impose strict constraints on the recommendation model $f_\Theta$, various alignment strategies can be flexibly applied. For discriminative methods, prevalent practice involves either directly minimizing the Euclidean distance between the two representations (\textit{e.g.}, mean squared error \cite{wang2020understanding}), or optimizing a lower bound of the mutual information $\mathbb I(\mathbf z_u, \tilde{\mathbf c}_{u}^e)$ to encourage semantic consistency across spaces (\textit{e.g.}, contrastive learning \cite{yu2022graph, ren2024representation}). For generative methods, we establish the mean $\tilde{\boldsymbol\mu}_{u}^{e}$ and variance $(\tilde{\boldsymbol\sigma}_{u}^{e})^2$ of the preference distribution for user $u$ in each environment $e$. Based on this, we focus primarily on aligning the learned distribution $\mathcal N(\tilde{\boldsymbol\mu}_{u}^{e}, (\tilde{\boldsymbol\sigma}_{u}^{e})^2)$ with the target distribution $\mathcal N(\boldsymbol{\mu}_u, \boldsymbol{\sigma}_u^2)$ (obtained by Eq. \ref{vae_q}), while incorporating regularization to prevent the posterior from diverging from the prior $p_{\text{prior}} \sim \mathcal N(\tilde{\boldsymbol{\mu}}^e_{\text{prior}}, (\tilde{\boldsymbol{\sigma}}_{\text{prior}}^e)^2)$ \cite{kingma2013auto, zhang2025towards}:
\begin{equation}
\label{KL_align}
\mathcal L_{\text{align}}^e= \sum_{o\in\{u, \text{prior}  \}}w_o\left (\text{ln}\frac{\tilde{\boldsymbol\mu}^e_o}{\boldsymbol\mu_{u}}+\frac{(\boldsymbol\sigma_u)^2+(\boldsymbol\mu_{u}-\tilde{\boldsymbol\mu}^e_o)^2}{2(\tilde{\boldsymbol\sigma}^e_o)^2} \right )-\frac{1}{2},
\end{equation}
where $w_{\text{prior}}=\beta, w_u= 1-\beta$, and $\beta \in [0, 1]$ is a trade-off coefficient. Eq. \ref{KL_align} essentially attempts to guide the alignment through the reverse KL divergence of the two input distributions \cite{zhang2025towards} (\textit{i.e.}, leveraging the original prior $p_{\text{prior}}$ and the distribution of each environment $e$ as the prior for the recommendation model $f_\Theta$, respectively). In practice, $p_{\text{prior}}$ is usually represented as a standard Gaussian $\mathcal N(\mathbf 0, \mathbf I)$ to facilitate optimization \cite{kingma2013auto}.

Once both dimension and semantic alignments are achieved, we integrate these representations to jointly perform the recommendation task.  Specifically, this requires the introduction of a fusion function $f_\gamma:\mathbb R^d \times \mathbb R^d 	\rightarrow \mathbb R^{d_\gamma}$:
\begin{equation}
\label{fuse}
\{\mathbf z_{u}^e\}_{\forall e \in \mathcal{E}} = \{f_\gamma(\mathbf z_u, \tilde{\mathbf c}_{u}^e),...|e\in \mathcal E\},
\end{equation}
where $d_\gamma$  denotes the dimensionality of the final representation. Benefiting from both dimension and semantic alignment, we empirically find that performing fusion with element-wise addition achieves satisfactory performance. Therefore, we adopt this simple yet effective design without introducing additional settings.

In each environment $e$, we utilize the fused representation $\mathbf z_{u}^e$ to perform the recommendation task, based on which the model-specific recommendation loss $\mathcal L^e_{\text{rec}}$ is computed. In addition, it is also necessary to consider the alignment process $\mathcal L^e_{\text{align}}$, which leads to the following multi-task objective:
\begin{equation}
\label{env_loss}
\mathcal L^e = \mathcal L^e_{\text{rec}}(\mathbf z_u^e) + \lambda_1 \cdot \mathcal L^e_{\text{align}}(\mathbf z_u, \tilde{\mathbf c}_{u}^e),
\end{equation}
where $\lambda_1$ is an adjustable weight. Due to the effects of random sampling and profile extrapolation, each environment $e$ exhibits significant variability. However, their essence still lies in characterizing the same user $u$, implying the existence of invariant features within such diversity. To ensure stable training and better capture the underlying user preferences, we impose a constraint on these environment-specific losses $\{\mathcal L^e\}_{\forall e \in \mathcal E}$ to reduce the discrepancy across different environments. Integrating the above components, we propose a model-agnostic recommendation framework \textsf{ProEx}, whose training objective is formulated as follows:
\begin{equation}
\label{loss}
\mathcal L_{\textsf{ProEx}}=\sum_{e\in \mathcal E}(\mathcal L^e)+\lambda_2 \cdot \mathcal L_{\text{reg}} + \lambda_3 \cdot \text{Var}_{e\in \mathcal E}(\mathcal L^e),
\end{equation}
where $\lambda_2$ and $\lambda_3$ are also adjustable weights. For each profile $k \in \mathcal K$, we apply contrastive regularization $\mathcal L_{\text{reg}}$ (\textit{cf.} Eq. \ref{reg}) to further enable profile extrapolation, thereby exploring a broader preference space. Finally, we minimize the variance across all environments $\mathcal E$ to reduce inter-environment discrepancies, thereby extracting the invariant components of user preferences and maintaining a more stable training process \cite{krueger2021out}. The complete training and inference 
processes of \textsf{ProEx} are presented in Algorithm \ref{algorithms} and Algorithm \ref{algorithms2}, respectively.

{\small
\begin{algorithm}
\caption{The inference process of \textsf{ProEx}}
\label{algorithms2}
\KwIn{user–item ineraction $\mathbf X$, base recommendation model $f_{\Theta}$, mapping function $f_\psi$, and fusion function $f_\gamma$.}
\begin{algorithmic}[1]
\STATE load model parameters for $f_{\Theta}$ and $f_\psi$;
\FOR{$u\in \mathcal U$}
\STATE construct recommendation representation $\mathbf z_u$ by $f_{\Theta}$;
\FOR{$k= 1, 2, ... \mathcal K$}
\STATE  transform the profile $s_{u,k}$ into semantic representation $\mathbf c_{u,k}$ by $f_{\text{text}}$ (Eq. \ref{trans});
\STATE construct dimension-aligned representation $\tilde{\mathbf c}_{u,k}$ by $f_{\psi}$ (Eq. \ref{align});
\ENDFOR

\STATE  construct mixed profile representation $\tilde{\mathbf c}_{u}$ by mean pooling;
\STATE  construct fused representation ${\mathbf z}_{u}$ by Eq. \ref{fuse};

\STATE compute interaction probability $p(\mathbf x_u|\mathbf z_u)$ for each item $i\in\mathcal I$ by Eq. \ref{score};
\ENDFOR

\RETURN Top-N recommendation list for each user $u \in \mathcal U$;
\end{algorithmic}
\end{algorithm}
}

\begin{table}[t]
\small
\setlength{\abovecaptionskip}{0.1cm}
\setlength{\belowcaptionskip}{0.1cm} 
  \caption{ Statistics of the datasets.}
  \label{dataset}
  \begin{tabular}{l|c|c|c|c}
    \hline
    \textbf{Dataset}&\textbf{\#Users}&\textbf{\#Items}&\textbf{\#Interactions}&\textbf{\#Profiles}\\
    \hline
    \hline
    \textbf{Amazon-Book}&11,000&9,332&200,860&81,328\\
    \textbf{Yelp}&11,091&11,010&277,535&88,404\\
    \textbf{Steam}&23,310&5,237&525,922&114,188\\
    \hline
  \end{tabular}
\end{table}

\begin{table*}
\small
        \centering
\setlength{\abovecaptionskip}{0.1cm}
\setlength{\belowcaptionskip}{0.1cm} 
  \caption{Overall performance comparisons between \textsf{ProEx} and base models on Amazon-Book, Yelp, and Steam datasets \textit{w.r.t.} Recall@N and NDCG@N (N $\in [10, 20]$). The best-performing model is highlighted in \textbf{bold}, and Improv.\% refers to the relative improvement of the best-performing model compared to the base model.}
  \label{performance1}
  \begin{tabular}{c|c|cccc|cccc|cccc}
    \hline
    \multicolumn{2}{c|}{\textbf{Model}}&\multicolumn{4}{c|}{\textbf{Amazon-Book}}&
    \multicolumn{4}{c|}{\textbf{Yelp}}&
    \multicolumn{4}{c}{\textbf{Steam}}\\
	\cline{1-14}		Base model&Variants&R@10&R@20&N@10&N@20&R@10&R@20&N@10&N@20&R@10&R@20&N@10&N@20\\
    \hline
    \hline
    \multicolumn{14}{c}{Discriminative Method}\\
    \hline
             \multirow{4}{*}{GCCF} &Base \cite{chen2020revisiting} & 0.0872 & 0.1343 & 0.0653 & 0.0807 & 0.0652 & 0.1084 & 0.0534 & 0.0680 & 0.0826 & 0.1313 & 0.0665 & 0.0830\\
    ~ &w/ \textsf{ProEx} & \textbf{0.0945} & \textbf{0.1443} & \textbf{0.0709} & \textbf{0.0873} & \textbf{0.0731} & \textbf{0.1201} & \textbf{0.0591} & \textbf{0.0751} & \textbf{0.0856} & \textbf{0.1361} & \textbf{0.0690} & \textbf{0.0861}\\
    \cline{2-14}
     ~ & Improv.\% & 8.37\% & 7.45\% & 8.58\% & 8.18\% & 12.12\% & 10.79\% & 10.67\% & 10.44\% & 3.63\% & 3.66\% & 3.76\% & 3.73\%\\
     ~ & $p$-values & 4.15e-6 & 7.11e-6 & 2.11e-5 & 1.27e-5 & 1.08e-4 & 3.25e-5 & 1.36e-4 & 8.75e-5 & 9.5e-7 & 3.31e-6 & 3.22e-7 & 1.79e-7\\
              \hline
    \multirow{4}{*}{LightGCN} &Base \cite{he2020lightgcn} & 0.0915 & 0.1411 & 0.0694 & 0.0856 & 0.0706 & 0.1157 & 0.0580 & 0.0733 & 0.0852 &0.1348 & 0.0687 & 0.0855\\
    ~ &w/ \textsf{ProEx} & \textbf{0.1008} & \textbf{0.1533} & \textbf{0.0770} & \textbf{0.0940} & \textbf{0.0800} & \textbf{0.1308} & \textbf{0.0653} & \textbf{0.0826} & \textbf{0.0937} & \textbf{0.1473} & \textbf{0.0756} & \textbf{0.0939}\\
    \cline{2-14}
     ~ & Improv.\% & 10.16\% & 8.65\% & 10.95\% & 9.81\% & 13.31\% & 13.05\% & 12.39\% & 12.30\% & 9.98\% & 9.27\% & 10.04\% & 9.82\%\\
     ~ & $p$-values & 2.19e-9 & 4.93e-9 & 9.2e-12 & 6.01e-12 & 5.46e-7 & 2.28e-8 & 6.63e-7 & 8.34e-8 & 1.41e-11 & 2.37e-11 & 4.95e-12 & 3.07e-12\\
              \hline
    \multirow{4}{*}{SimGCL} &Base \cite{yu2022graph} & 0.0992 & 0.1512 & 0.0749 & 0.0919 & 0.0772 & 0.1254 & 0.0638 & 0.0801 & 0.0918 &0.1436 & 0.0738 & 0.0915\\
    ~ &w/ \textsf{ProEx} & \textbf{0.1041} & \textbf{0.1574} & \textbf{0.0794} & \textbf{0.0967} & \textbf{0.0821} & \textbf{0.1331} & \textbf{0.0681} & \textbf{0.0855} & \textbf{0.0948} & \textbf{0.1483} & \textbf{0.0767} & \textbf{0.0947}\\
    \cline{2-14}
     ~ & Improv.\% & 4.83\% & 3.92\% & 5.67\% & 4.99\% & 8.26\% & 7.03\% & 8.87\% & 8.19\% & 3.27\% & 3.27\% & 3.93\% & 3.50\%\\
     ~ & $p$-values & 5.76e-6 & 6.84e-6 & 7.8e-8 & 8.09e-8 & 1.88e-8 & 2.37e-8 & 1.35e-9 & 1.55e-9 & 1.67e-4 & 4.67e-6 & 1.67e-7 & 1.06e-7 \\
              \hline
              \multicolumn{14}{c}{Generative Method}\\
              \hline
    \multirow{4}{*}{Mult-VAE} &Base \cite{liang2018variational} & 0.0976 & 0.1472 & 0.0752 & 0.0916 & 0.0732 & 0.1189 & 0.0593 & 0.0751 & 0.0929 &0.1452 & 0.0744 & 0.0923\\
    ~ & \textsf{w/ ProEx} & \textbf{0.1091} & \textbf{0.1596} & \textbf{0.0824} & \textbf{0.0997} & \textbf{0.0795} & \textbf{0.1289} & \textbf{0.0644} & \textbf{0.0813} & \textbf{0.1003} & \textbf{0.1548} & \textbf{0.0799} & \textbf{0.0987}\\
    \cline{2-14}
     ~ & Improv.\% & 11.78\% & 8.42\% & 9.57\% & 8.84\% & 8.61\% & 8.41\% & 8.60\% & 8.26\% & 7.97\% & 6.61\% & 7.39\% & 6.93\%\\
~ & $p$-values & 5.82e-7 & 5.62e-8 & 2.62e-6 & 6.67e-7 & 8.52e-4 & 3.03e-7 & 6.88e-5 & 9.44e-7 & 6.24e-9 & 1.45e-9 & 8.17e-9 & 7.19e-10\\
      \hline
        \multirow{4}{*}{L-DiffRec} &Base \cite{wang2023diffusion} & 0.1048 & 0.1495 & 0.0844 & 0.0990 & 0.0721 & 0.1177 & 0.0590 & 0.0745 &0.0885 & 0.1395 & 0.0722 & 0.0893\\
    ~ &w/ \textsf{ProEx} & \textbf{0.1127} & \textbf{0.1593} & \textbf{0.0901} & \textbf{0.1054} & \textbf{0.0772} & \textbf{0.1254} & \textbf{0.0642} & \textbf{0.0805} & \textbf{0.0971}& \textbf{0.1494} & \textbf{0.0785} & \textbf{0.0965}\\
    \cline{2-14}
     ~ & Improv.\% & 7.54\% & 6.56\% & 6.75\% & 6.46\% & 7.07\% & 6.54\% & 8.81\% & 8.05\% & 9.72\% & 7.10\% & 8.73\% & 8.06\%\\
     ~ & $p$-values & 6.09e-4 & 5.59e-5 & 7.24e-5 & 3.25e-6 & 2.4e-5 & 1.95e-5 & 9.72e-7 & 2.21e-7 & 3.35e-5 & 3.43e-5 & 2.36e-4 & 9.43e-5\\
         \hline
    \multirow{4}{*}{CVGA} &Base \cite{zhang2023revisiting} & 0.1030 & 0.1522 & 0.0799 & 0.0960 & 0.0779 & 0.1249 & 0.0639 & 0.0801 &0.0842 &0.1339 & 0.0679 & 0.0847\\
    ~ &w/ \textsf{ProEx} & \textbf{0.1140} & \textbf{0.1660} & \textbf{0.0880} & \textbf{0.1050} & \textbf{0.0825} & \textbf{0.1339} & \textbf{0.0672} & \textbf{0.0847} & \textbf{0.0998} & \textbf{0.1554} & \textbf{0.0801} & \textbf{0.0991} \\
    \cline{2-14}
     ~ & Improv.\% & 9.82\% & 9.07\% & 10.16\% & 9.41\% & 5.94\% & 7.17\% & 5.09\% & 5.71\% & 18.54\% & 16.03\% & 17.91\% & 17.04\%\\
     ~ & $p$-values & 7.03e-8 & 2.4e-9 & 3.19e-8 & 1.62e-9 & 9.35e-6 & 2.85e-10 & 3.13e-8 & 707e-10 & 2.27e-13 & 1.25e-14 & 1.91e-14 & 1.76e-15\\

\hline
  \end{tabular}
\end{table*}

\section{Experiments}
In this section, to demonstrate the capability of \textsf{ProEx}, we carry out detailed experiments and analyses.
\subsection{Experiment Settings}
\subsubsection{\textbf{Datasets}} We utilize three widely adopted recommendation datasets — Amazon-Book, Yelp, and Steam \cite{ren2024representation}. Detailed statistics for three datasets are presented in Table \ref{dataset}. Each dataset is divided into training, validation, and test sets following a 3:1:1 split ratio. The last column shows the number of user and item profiles for each dataset, where each user (item) has one original profile (OP) and three new profiles (NP \#1, NP \#2, and NP \#3).

\subsubsection{\textbf{Base Recommender}}
\label{base_model}
For the proposed \textsf{ProEx}, we select the following classic recommenders as the base models:

\noindent
\textbf{discriminative method}: GCCF \cite{chen2020revisiting} , LightGCN \cite{he2020lightgcn}, SimGCL \cite{yu2022graph};

\noindent
\textbf{generative method}: Mult-VAE \cite{liang2018variational}, DiffRec \cite{wang2023diffusion}, CVGA \cite{zhang2023revisiting}.

\subsubsection{\textbf{LLM-enhanced Baselines}}
To enable a more thorough comparison, we additionally include the following LLM-enhanced methods as baselines:

\noindent
    \textbf{discriminative method}: CARec \cite{wang2024collaborative}, KAR \cite{xi2024towards}, LLMRec \cite{wei2024llmrec}, RLMRec\cite{ren2024representation}, AlphaRec \cite{sheng2025language};

    \noindent
    \textbf{generative method}: DMRec \cite{zhang2025towards}.

\subsubsection{\textbf{Implementation Details}} We implement \textsf{ProEx} by PyTorch\footnote{https://github.com/BlueGhostYi/ProRec}. All models are initialized using Xavier initialization \cite{glorot2010understanding} and optimized with the Adam optimizer \cite{kingma2014adam}. The architectures of the base models adhere to the default configurations. The default batch size is 4,096 for discriminative methods, in contrast to 1,024 for generative methods. For all LLM-enhanced methods, we use OpenAI's GPT-3.5-turbo as the base LLM and text-embedding-ada-002 \cite{neelakantan2022text} for text embeddings to ensure a fair comparison. To evaluate the recommendation performance, we use the metrics Recall@N and NDCG@N \cite{he2017neural}. The full-ranking evaluation \cite{he2020lightgcn} is applied for each user. Early stopping is activated when Recall@20 on the validation set does not improve for 20 consecutive epochs. All experiments are conducted over 10 random seeds, with the averaged results reported. Additional details can be found in Appendix \ref{settings}.

\subsection{Performance Comparisons}
\subsubsection{\textbf{Model-agnostic Performance Comparison}}
\begin{figure}
\setlength{\abovecaptionskip}{0.0cm}
\setlength{\belowcaptionskip}{0.0cm} 
\centering
\subfigure[Amazon-Book]{\includegraphics[width=0.15\textwidth]{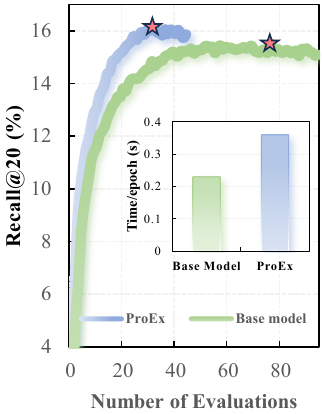}
\label{amazon_time_case}}
\hfil
\subfigure[Yelp]{\includegraphics[width=0.15\textwidth]{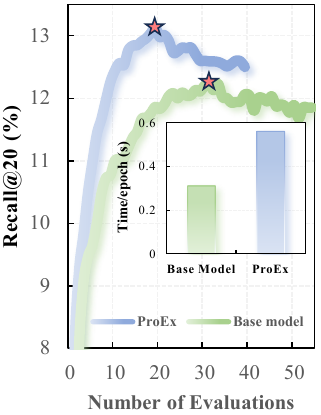}
\label{yelp_time_case}}
\hfil
\subfigure[Steam]{\includegraphics[width=0.15\textwidth]{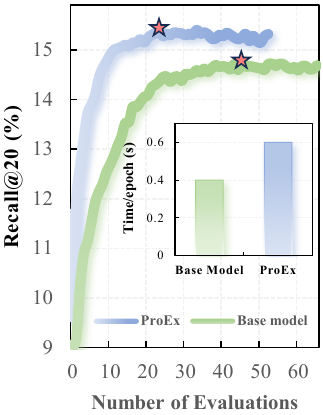}
\label{steam_time_case}}

\caption{Comparison of the training process and time cost of the base model and \textsf{ProEx} \textit{w.r.t.} Recall@20 on validation sets. The red star indicates the best-performing on test sets.}
\label{fig_time}
\end{figure}

To evaluate the generalization capability of \textsf{ProEx}, we integrate it with the three discriminative methods and three generative methods introduced in Section \ref{base_model}. The experimental results are presented in Table \ref{performance1}, from which we derive the following observations.
\begin{itemize}[leftmargin=*]
\item Intuitively, \textsf{ProEx} achieves the best performance over all base models on three datasets. Quantitatively, taking LightGCN as an example, \textsf{ProEx} improves by 8.67\%, 13.05\% and 9.11\% over the base model \textit{w.r.t.} Recall@20 on Amazon-Book, Yelp, and Steam datasets, respectively. The above experimental results demonstrate the effectiveness of the proposed \textsf{ProEx}.

\item In general, \textsf{ProEx} achieves a significant improvement of more than 5\% on average for six base models, demonstrating strong generalizability across both discriminative and generative methods. It is worth noting that the improvement on generative methods is more stable. One possible reason is that the encoders of generative methods usually have larger dimensions (\textit{e.g.}, 200), which are closer to the dimensionality of the semantic vectors produced by LLMs. In contrast, the embedding dimensions of discriminative methods are typically much smaller (\textit{e.g.}, 32), so the dimension alignment process may cause more semantic loss.

\item Finally, Mult-VAE is adopted as a case study to compare the training efficiency between the base model and \textsf{ProEx}, as illustrated in Fig. \ref{fig_time}. \textsf{ProEx} substantially enhances the training efficiency, with only a minor increase in time per iteration.
\end{itemize}

\subsubsection{\textbf{Performance Comparison with LLM-enhanced Methods}}
To further demonstrate the effectiveness of \textsf{ProEx}, we conduct comparison experiments with other LLM-enhanced recommendation methods, starting with discriminative methods. For a fair comparison, we use LightGCN \cite{he2020lightgcn} as the base model, set the embedding size uniformly to 32 \cite{ren2024representation}, and employ the same LLM and text embedding model. The comparison results\footnote{AlphaRec* is a variant of AlphaRec \cite{sheng2025language} additionally utilizes user-side representations learned from LLMs.} between \textsf{ProEx} and other baselines are shown in Table \ref{tab:LLM1}. Intuitively, \textsf{ProEx} achieves the best recommendation performance on both metrics across all datasets. Although all the selected LLM-enhanced methods use profiles or semantic representations derived from the LLM, \textsf{ProEx} can fully leverage multiple profiles to better capture user preferences.

For generative methods, we adopt Mult-VAE \cite{liang2018variational} as the base model to compare \textsf{ProEx} with the recent work DMRec \cite{zhang2025towards}. In addition, we apply each profile individually to DMRec-M and report the average result over all profiles (denoted as $\text{DMRec}_{\text{avg}}$). The results are shown in Table \ref{tab:LLM2}. It can be observed that \textsf{ProEx} still achieves further performance improvement. Moreover, $\text{DMRec}_{\text{avg}}$ exhibits considerable performance variations compared to DMRec. This not only indicates that relying on single profile inevitably introduces bias, but also demonstrates that \textsf{ProEx} can effectively leverage multiple profiles to avoid such single-profile dependency.

\begin{table}[!t]

\setlength{\abovecaptionskip}{0.1cm}
\setlength{\belowcaptionskip}{0.1cm} 
\centering
\small
  \caption{Comparison of recommendation performance between \textsf{ProEx} and LLM-enhanced recommendation methods based on LightGCN \cite{he2020lightgcn} \textit{w.r.t.} Recall@20 and NDCG@20.}
  \label{tab:LLM1}
  \begin{tabular}{l|cc|cc|cc}
    \hline
        &\multicolumn{2}{c|}{\textbf{Amazon-Book}}&
    \multicolumn{2}{c|}{\textbf{Yelp}}&\multicolumn{2}{c}{\textbf{Steam}}\\
	\cline{2-7}	
    &R@20&N@20&R@20&N@20&R@20&N@20\\
    \hline
    \hline
CARec&0.1391&0.0854&0.1130&0.0714&0.1384& 0.0880\\
\hline
KAR&0.1416&0.0863&0.1194&0.0756&0.1353& 0.0854\\
    \hline
LLMRec&0.1469&0.0855&0.1203&0.0751&0.1431&0.0901\\
    \hline
RLMRec-C&\underline{0.1483}&\underline{0.0903}&\underline{0.1230}&\underline{0.0776}&0.1421&0.0902\\
RLMRec-G&0.1446&0.0887&0.1209&0.0761&\underline{0.1433}&\underline{0.0907}\\
\hline
AlphaRec&0.1412&0.0873&0.1212&0.0752&0.1404&0.0889\\
AlphaRec*&0.1421&0.0835&0.1213&0.0752&0.1420&0.0898\\
\hline
\hline
\textbf{\textsf{ProEx}}&\textbf{0.1533}&\textbf{0.0940}&\textbf{0.1308}&\textbf{0.0826}&\textbf{0.1473}&\textbf{0.0939}\\
   \hline
\end{tabular}
\end{table}

\begin{table}[!t]

\setlength{\abovecaptionskip}{0.1cm}
\setlength{\belowcaptionskip}{0.1cm} 
\centering
\small
  \caption{Comparison of recommendation performance between \textsf{ProEx} and DMRec based on  Mult-VAE \cite{liang2018variational} \textit{w.r.t.} Recall@20 and NDCG@20.}
  \label{tab:LLM2}
  \begin{tabular}{l|cc|cc|cc}
    \hline
        &\multicolumn{2}{c|}{\textbf{Amazon-Book}}&
    \multicolumn{2}{c|}{\textbf{Yelp}}&\multicolumn{2}{c}{\textbf{Steam}}\\
	\cline{2-7}	
    &R@20&N@20&R@20&N@20&R@20&N@20\\
    \hline
    \hline
DMRec-G&0.1545&0.0960&0.1226&0.0775&0.1495& 0.0950\\
DMRec-C&0.1561&0.0971&0.1254&\underline{0.0792}&0.1496& 0.0948\\
DMRec-M&\underline{0.1571}&\underline{0.0979}&\underline{0.1261}&0.0785&\underline{0.1536}& \underline{0.0973}\\
\hline
\hline
\textbf{\textsf{ProEx}}&\textbf{0.1596}&\textbf{0.0997}&\textbf{0.1289}&\textbf{0.0813}&\textbf{0.1548}&\textbf{0.0987}\\
   \hline
\end{tabular}
\end{table}

\begin{figure}
\setlength{\abovecaptionskip}{0.0cm}
\setlength{\belowcaptionskip}{0.0cm} 
\centering
\subfigure[LightGCN]{\includegraphics[width=1.62in]{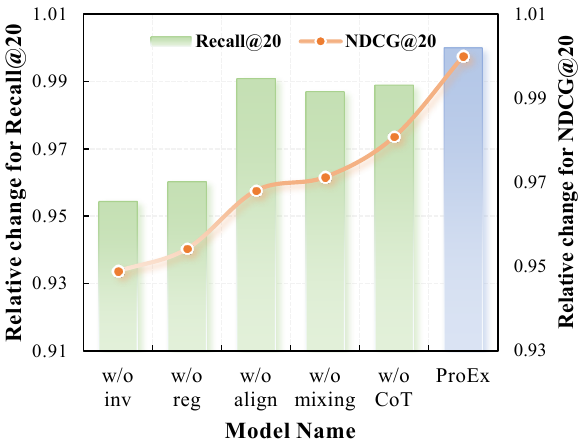}
\label{ablation_case_1}}
\hfil
\subfigure[Mult-VAE]{\includegraphics[width=1.62in]{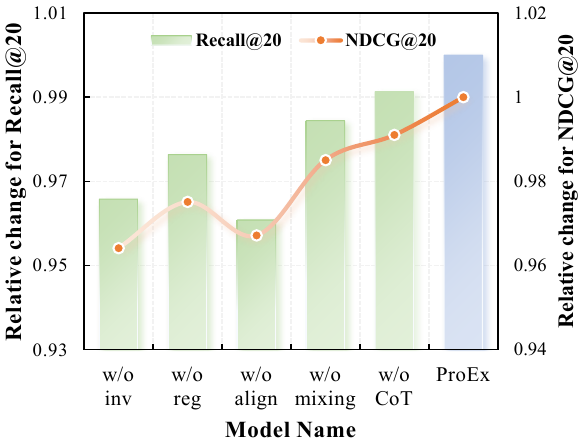}
\label{ablation_case_2}}

\caption{Ablation studies for (a) LightGCN and (b) Mult-VAE \textit{w.r.t.} Recall@20 and NDCG@20 on Amazon-Book dataset.}
\label{fig_ablation}
\end{figure}

\subsection{In-depth Analysis of ProEx}
\subsubsection{\textbf{Ablation Studies}}
In this section, we construct several variants to validate the necessity of some of \textsf{ProEx}'s design choices:
\begin{itemize}[leftmargin=*]
\item $\textsf{ProEx}_{\text{w/o inv}}$: remove the variance term in Eq. \ref{loss} and directly use the first environment for recommendation;
\item $\textsf{ProEx}_{\text{w/o reg}}$: remove the regularization loss $\mathcal L_{\text{reg}}$ in Eq. \ref{loss};
\item $\textsf{ProEx}_{\text{w/o align}}$: remove the cross-space alignment loss in each environment $e$ in Eq. \ref{env_loss};
\item $\textsf{ProEx}_{\text{w/o mixing}}$: remove the process of constructing profiles in Eqs. \ref{mix_1} and \ref{mix_2}, and instead use average pooling;
\item $\textsf{ProEx}_{\text{w/o CoT}}$: remove the four-step chain-of-thought reasoning in Eq. \ref{CoT} and directly have the LLM generate multiple new profiles.
\end{itemize}
The experimental results for \textsf{ProEx} and all variants on LightGCN and Mult-VAE are shown in Fig. \ref{fig_ablation}. It is evident that all variants achieve lower performance than \textsf{ProEx}, thereby demonstrating the validity and effectiveness of the proposed components. Overall, $\textsf{ProEx}_{\text{w/o inv}}$ and $\textsf{ProEx}_{\text{w/o reg}}$ lead to a significant performance degradation, highlighting the importance of profile extrapolation. The impact of cross-space alignment varies between the two base models, which is primarily determined by the characteristics of the models themselves. Finally, although the introduction of multiple environments and CoT reasoning slightly increases the training time per iteration, it leads to further performance gains for \textsf{ProEx}, thereby justifying the added complexity.

\begin{figure}[t]
\setlength{\abovecaptionskip}{0.0cm}
\setlength{\belowcaptionskip}{0.0cm} 
  \centering
  \includegraphics[width=\linewidth]{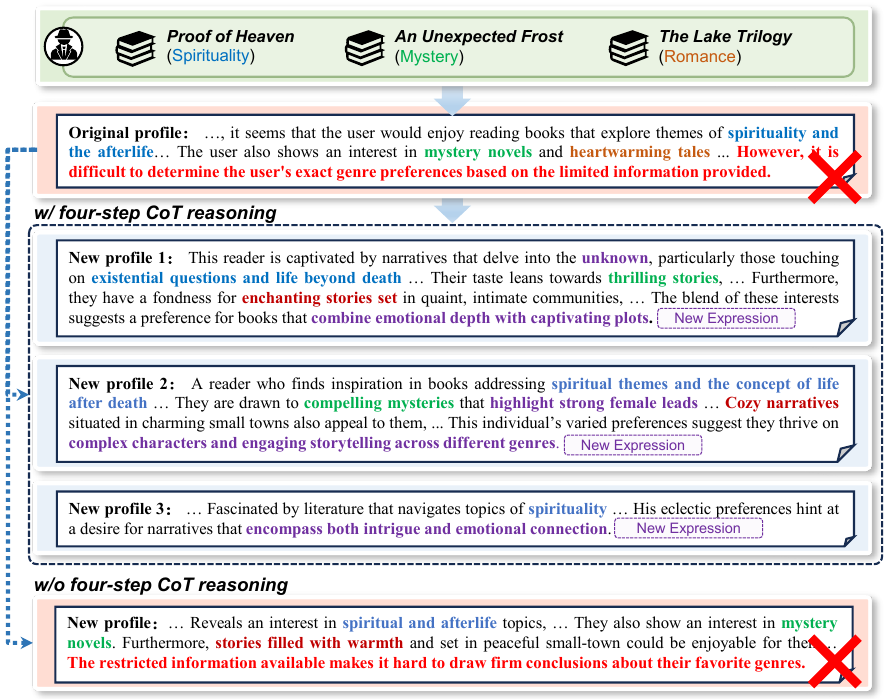}
  \caption{Comparison between the original and the new profiles generated w/ and w/o four-step CoT reasoning for user \#3638. Identical colors denote the same type of expression.}
\label{fig_case_study}
\end{figure}

For the CoT reasoning, we randomly select user \#3568 from Amazon-Book dataset, whose original profile is non-sensical, and we compare the new profiles generated with and without the application of four-step CoT reasoning, as shown in Fig. \ref{fig_case_study}. It can be observed that when only given a prompt to regenerate the profile, the LLM merely attempts to modify the wording and structure of the original profile, yet still fails to capture the user’s preferences. In contrast, the new profiles generated via CoT reasoning not only greatly enrich the original descriptions but also introduce a substantial amount of new information (in purple color).

\begin{table}[!t]

\setlength{\abovecaptionskip}{0.1cm}
\setlength{\belowcaptionskip}{0.1cm} 
\centering
\small
  \caption{Comparison of recommendation performance with different semantic representations on LightGCN and Mult-VAE \textit{w.r.t.} Recall@20 and NDCG@20.}
  \label{tab:LLM}
  \begin{tabular}{l|cc|cc}
    \hline
        &\multicolumn{2}{c|}{\textbf{LightGCN}}&
    \multicolumn{2}{c}{\textbf{Mult-VAE}}\\
	\cline{2-5}	
    &R@20&N@20&R@20&N@20\\
    \hline
    \hline
base&0.1411&0.0856&0.1472&0.0916\\
\hline
w/ Llama3-8B-Instruct \cite{dubey2024llama}&0.1431&0.0870&0.1547&0.0954\\
w/ text-embedding-ada-002 \cite{neelakantan2022text}&0.1533&0.0940&0.1596&0.0997\\
w/ text-embeddings-3-large \cite{neelakantan2022text}&0.1466&0.0893&0.1625&0.1014\\
w/ SFR-Embedding-Mistral \cite{meng2024sfrembedding}&0.1515&0.0923&0.1580 & 0.0978\\
   \hline
\end{tabular}
\end{table}

\subsubsection{\textbf{Impact of Different Semantic Representations}} Transforming the generated profiles into representations in the feature space is a crucial step in \textsf{ProEx}, which is primarily accomplished through the text embedding model $f_{\text{text}}$. We conduct experiments using various semantic representations on the base models LightGCN \cite{he2020lightgcn} and Mult-VAE \cite{liang2018variational}, as shown in Table \ref{tab:LLM}. 

Firstly, all types of semantic representations improve the performance of the base models to varying degrees, further confirming the effectiveness and generalizability of \textsf{ProEx}.
Secondly, although LLaMA3 produces the largest embeddings (4096), it achieves the worst performance, particularly with LightGCN. This may stem from the limited semantic understanding of the 8B-scale model, leading to low-quality representations. Moreover, the severe dimensional mismatch (e.g., 32 dimensions in LightGCN) makes alignment more difficult. Finally, the locally deployable SFR-Embedding-Mistral achieves strong performance, suggesting that \textsf{ProEx} does not necessarily rely on closed-source LLMs to be effective.

\subsubsection{\textbf{Hyper-parameter Sensitivities}}

\begin{figure}
\setlength{\abovecaptionskip}{0.0cm}
\setlength{\belowcaptionskip}{0.0cm} 
\centering
\subfigure[LightGCN]{\includegraphics[width=1.62in]{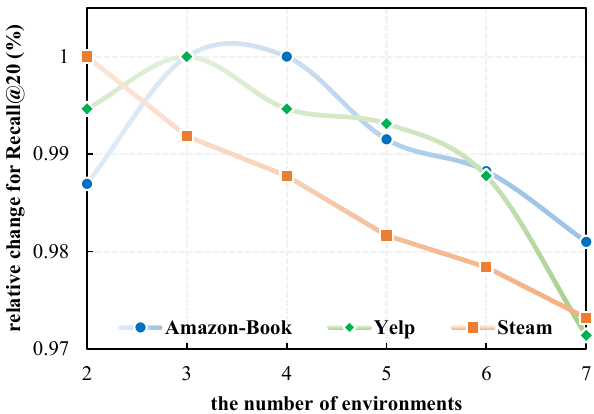}
\label{env_lightgcn_case}}
\hfil
\subfigure[Mult-VAE]{\includegraphics[width=1.62in]{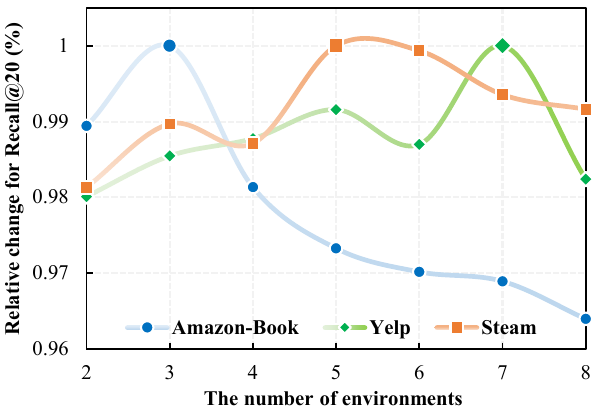}
\label{env_vae_case}}

\caption{Impact of the number of environments $|\mathcal E|$ on (a) LightGCN and (b) Mult-VAE \textit{w.r.t.} Recall@20.}
\label{fig_env}
\end{figure}

\begin{figure}
\setlength{\abovecaptionskip}{0.0cm}
\setlength{\belowcaptionskip}{0.0cm} 
\centering
\subfigure[LightGCN]{\includegraphics[width=1.62in]{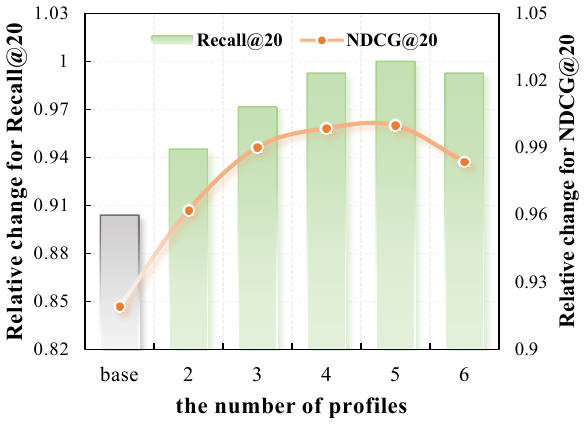}
\label{pro_lightgcn_case}}
\hfil
\subfigure[Mult-VAE]{\includegraphics[width=1.62in]{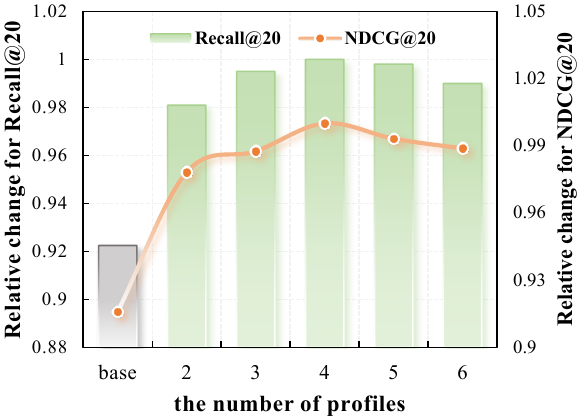}
\label{pro_vae_case}}

\caption{Impact of $K$ for (a) LightGCN and (b) Mult-VAE \textit{w.r.t.} Recall@20 and NDCG@20 on Amazon-Book dataset.}
\label{fig_pro}
\end{figure}

In this section, we first investigate the impact of two important hyperparameters, $|\mathcal E|$ and $K$, as shown in Figs. \ref{fig_env} and \ref{fig_pro}, respectively. For the number of environments $|\mathcal E|$, LightGCN performs better with fewer environments (2–3), while Mult-VAE can accommodate a larger number of environments. For the number of profiles $K$, both methods achieve the best performance with four or five profiles. An excessive number of profiles may not only introduce additional noise into the training process but also incur additional training costs.

\begin{figure}
\setlength{\abovecaptionskip}{0.0cm}
\setlength{\belowcaptionskip}{0.0cm} 
\centering
\subfigure[LightGCN]{\includegraphics[width=1.62in]{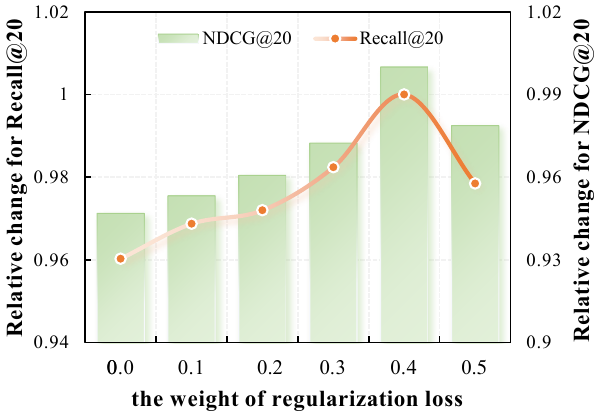}
\label{lambda_gcn_case}}
\hfil
\subfigure[Mult-VAE]{\includegraphics[width=1.62in]{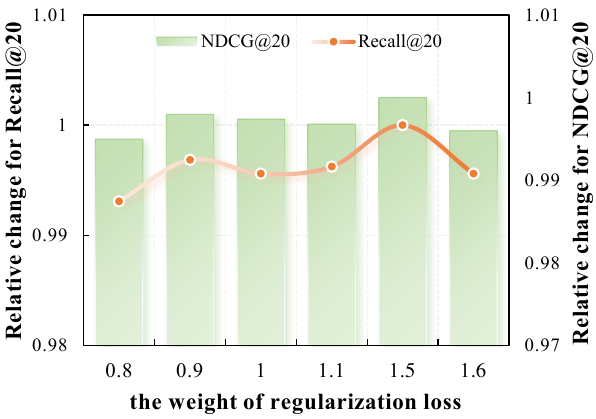}
\label{lambda_vae_case}}

\caption{Impact of $\lambda_2 $ for (a) LightGCN and (b) Mult-VAE \textit{w.r.t.} Recall@20 and NDCG@20 on Amazon-Book dataset.}
\label{fig_lambda}
\end{figure}

Finally, we investigate the weight of the contrastive regularization loss $\lambda_2$ used to enhance profile extrapolation, as shown in Fig. \ref{fig_lambda}. It can be seen that Mult-VAE requires a larger loss weight, while LightGCN only needs a small range. This is expected: LightGCN uses a pairwise loss \cite{rendle2009bpr} that is typically below 1, whereas Mult-VAE relies on a much larger reconstruction error (often tens to hundreds). Thus, a higher weight is needed to place the two losses on a comparable scale.

\section{Related Work}
Owing to the strong text processing capabilities of large language models \cite{zhao2023survey}, their integration into recommender systems has attracted increasing research interest \cite{geng2022recommendation, bao2023tallrec}. Existing approaches can be broadly categorized into two paradigms. The first line of research typically integrates the recommendation task into the fine-tuning process of LLMs \cite{bao2023tallrec, touvron2023llama}. For example, P5 \cite{geng2022recommendation} directly converts user–item interaction data into textual prompts for training. Building on this idea, subsequent works such as TallRec \cite{bao2023tallrec} and LLaRA \cite{liao2024llara} introduce adapters or leverage LoRA \cite{hulora2022} to enable more parameter-efficient fine-tuning. However, despite these advancements, fine-tuning approaches generally demand substantial computational resources and training time \cite{ren2024representation}, and their applicability is often limited by specific deployment scenarios \cite{liao2024llara}. Specifically, these fine-tuning-based methods are typically applied to sequential or session recommendation and struggle to achieve satisfactory performance in general recommendation tasks \cite{geng2022recommendation} based on full ranking \cite{he2020lightgcn}, suggesting that solely relying on LLMs for recommendation or re-ranking may still be insufficient \cite{ren2024representation, sheng2025language}. 

The second line of research preserves the base recommendation model while incorporating the LLM as an auxiliary component. For example, multiple agents can be constructed by leveraging the reasoning and memory capabilities of LLMs to assist in making recommendation decisions \cite{zhang2024generative}, or the text processing and generation capabilities of LLMs can be utilized to create richer representations of users and items \cite{ren2024representation}. This type of method emphasizes primarily focus on achieving consistency within the feature space, where the key objective is to transform textual prompts into actionable latent representations \cite{xi2024towards, zhao2025continual}. Works such as KAR \cite{xi2024towards}, RLMRec \cite{ren2024representation}, and DMRec \cite{zhang2025towards}, highlight the importance of aligning the representations of recommendation models with those of LLMs \cite{radford2021learning}. However, given the sparsity of recommendation data and the inherent instability of LLMs, the textual information (\textit{i.e.}, profiles \cite{wang2025lettingo}) inferred from the original data may contain errors or biases, which can interfere with downstream recommendation.

\section{Conclusion}
In this paper, we revisited LLM-based profiling for recommendation and proposed a unified recommendation framework called \textsf{ProEx}. The core of \textsf{ProEx} lies in generating multi-faceted profiles for each user and item through a four-step chain-of-thought reasoning process. After performing dimension and semantic alignment, we introduced the concept of environments to extract the invariance of user preferences from multiple profiles. Empirical experiments on three datasets demonstrated that \textsf{ProEx} could improve the performance of various types of base recommendation models, while also showing advantages over other LLM-enhanced methods.

\begin{acks}
This work is supported by the National Natural Science Foundation of China (No. 62272001), the Australian Research Council under the streams of Future Fellowship (No. FT210100624), the Discovery Early Career Researcher Award (No. DE230101033), the Discovery Project (No. DP240101108, DP240101814, and DP260100326), and the Linkage Projects (No. LP230200892 and LP240200546). 
\end{acks}

\bibliographystyle{ACM-Reference-Format}
\bibliography{sample-base}

\newpage
\appendix

\balance
\section{Appendix}
In the Appendix, we provide a detailed introduction to the prompt designs of
multi-faceted profile construction, all comparation methods and hyperparameter settings.

\begin{figure}[hb]
\setlength{\abovecaptionskip}{0.1cm}
\setlength{\belowcaptionskip}{0.1cm} 
  \centering
  \includegraphics[width=\linewidth]{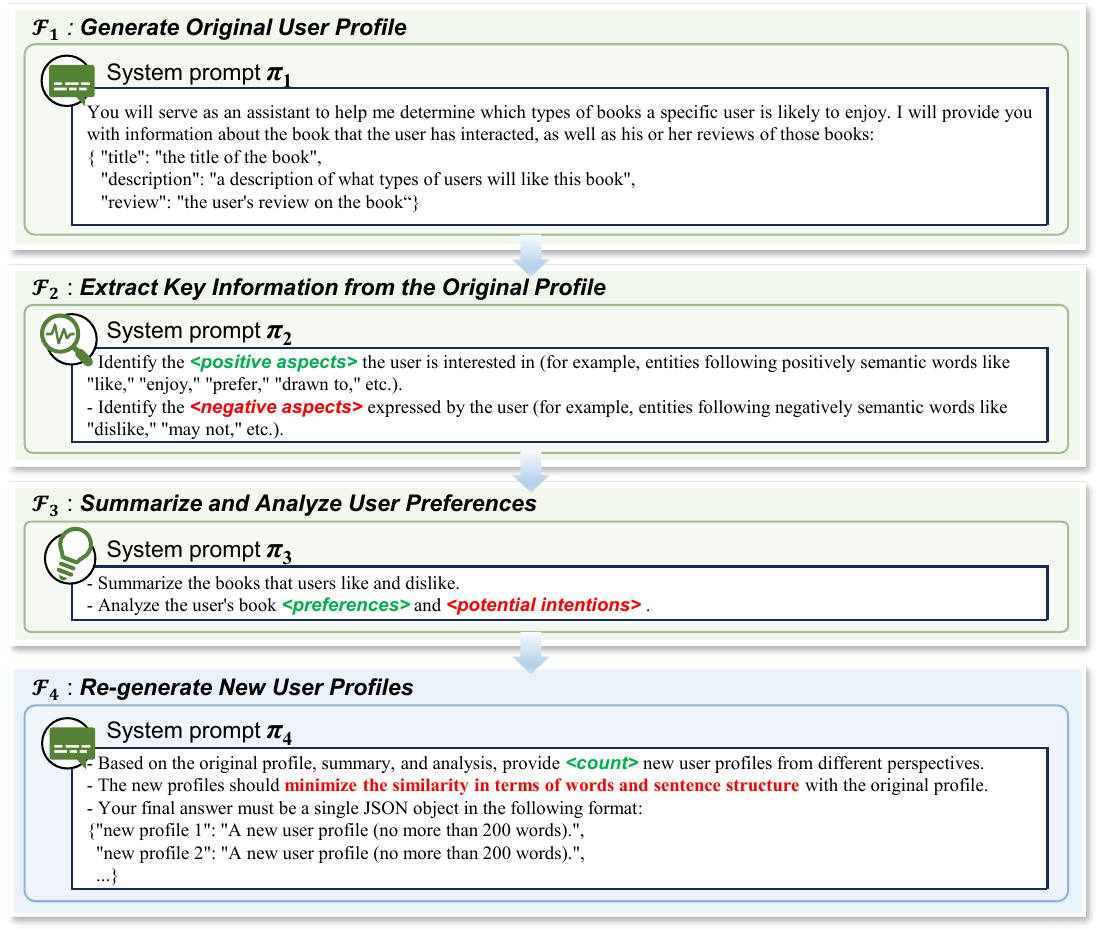}
  \caption{The detailed multi-faceted profile construction process based on four-step chain-of-thought reasoning for user side on Amazon-Book dataset.}
    \label{fig_cot2}
\end{figure}

\subsection{Multi-faceted Profile Construction}
The detailed process of generating multi-faceted profiles, along with the corresponding prompt design, is illustrated in Fig. \ref{fig_cot2} for the Amazon-Book dataset. A similar definition is applied on the item side and other datasets. To ensure consistency, the design of $\mathcal F_1$ follows prior work \cite{ren2024representation}. We input the textual information associated with the user $u$ (title, description, and review of interacted item) to obtain the original profile. In $\mathcal F_2$, information from the OP is extracted to identify the user $u$’s positive and negative aspects toward interacted items. Subsequently, in $\mathcal F_3$, the user $u$’s latent preferences are further analyzed. All the above information is then summarized and used in $\mathcal F_4$ to generate multiple new profiles. To encourage diversity and mitigate bias, we impose a constraint that the newly generated profiles exhibit maximal differences in lexical choices and syntactic structures. In practice, the original profile from $\mathcal F_1$ may contain incorrect or even meaningless statements (e.g., the original profile shown in Fig. \ref{fig_case_study}). Therefore, we perform more in-depth analysis of the original profile in $\mathcal F_2$ and $\mathcal F_3$ to extract preference-relevant information as much as possible from the limited content. In $\mathcal F_4$, we rephrase the statements to reduce reliance on a single profile.

\subsection{Experimental Details}
\label{exp_section}

\subsubsection{\textbf{Base Model Introduction}}
For the proposed \textsf{ProEx}, we select the following classic recommenders as the base models.

Discriminative Methods:
\begin{itemize}[leftmargin=*]
    \item \textbf{GCCF} \cite{chen2020revisiting}: This method employs the standard graph convolution operation to model the high-order connectivity between users and items, and obtains the final user and item embeddings through concatenation.
    \item \textbf{LightGCN} \cite{he2020lightgcn}:This method significantly simplifies the design of standard graph convolution by constructing user and item embeddings through linear aggregation of neighbors, and derives the final embedding representation via average pooling.
    \item \textbf{SimGCL} \cite{yu2022graph}:This method introduces contrastive learning on top of LightGCN, performing data augmentation by injecting additional noise into the embeddings of users and items.
\end{itemize}

Generative Methods:
\begin{itemize}[leftmargin=*]
    \item \textbf{Mult-VAE} \cite{liang2018variational}: This method is a standard generative recommendation approach based on variational auto-encoders, modeling user preferences through a non-linear network.
    \item \textbf{L-DiffRec} \cite{he2020lightgcn}:This method explores the use of diffusion models in generative recommendation by progressively adding Gaussian noise in the distribution space instead of directly perturbing the original data.
    \item \textbf{CVGA} \cite{zhang2023revisiting}:This method builds upon the basic variational auto-encoder by incorporating graph structure modeling, with an overall design similar to that of a variational graph autoencoder.
\end{itemize}

\subsubsection{\textbf{LLM-enhanced Baselines}}
To enable a more thorough comparison, we additionally include the following LLM-enhanced methods as baselines.
\begin{itemize}[leftmargin=*]
    \item \textbf{CARec} \cite{wang2024collaborative} integrates collaborative filtering signals into the semantic representations produced by LLMs through collaborative alignment.
    \item \textbf{KAR} \cite{xi2024towards} creates textual profiles for users and items, and integrates the LM-enhanced representations with recommenders through a hybrid-expert adaptor.
    \item \textbf{LLMRec} \cite{wei2024llmrec} enhances data reliability by employing graph augmentation strategies based on language models and a denoising mechanism for data robustification.
    \item \textbf{RLMRec} \cite{ren2024representation} aligns semantic representations of users and items between the LM-enhanced and recommendation representations. The contrastive strategy is referred to as RLMRec-C, while the generative strategy is referred to as RLMRec-G.
    \item \textbf{AlphaRec} \cite{sheng2025language} is a recently proposed LLM-based recommendation method that directly applies non-linear mapping and graph convolution operations on LLM-enhanced item representations.
    \item \textbf{DMRec} \cite{zhang2025towards} is an LLM-enhanced method specifically designed for generative models. It models user profiles and enhances the recommendation process through distribution matching. The global optimality strategy is referred to as DMRec-G, the composite prior strategy is referred to as DMRec-C, and the mixing divergence strategy is referred to as DMRec-M.
\end{itemize}

\subsubsection{\textbf{Detailed Hyperparameter Settings}}
\label{settings}
The architectures of the baseline methods adhere to the default configurations as specified in their respective original works. For discriminative methods, we set the learning rate to 0.001, the training batch size to 4096, the embedding dimension to 32, and the number of GCN layers to 3. For SimGCL \cite{yu2022graph}, contrastive learning-related hyperparameters are tuned separately for each dataset. Specifically, the temperature coefficient is searched over the range 
[0.2, 0.3, 0.4, 0.5], the weight of the contrastive loss is selected from [0.01, 0.1, 0.2, 0.5, 1.0], and the noise strength is searched within [0.1,1.0] with a step size of 0.1.

For generative methods, we follow the model architecture suggested in the original paper. Specifically, for Mult-VAE \cite{liang2018variational}, the architecture is set to [200,600] (\textit{i.e.}, the dimensionality of the user preference distribution is 200), the decay coefficient is in [0.1, 0.9] with a step size of 0.1. For L-DiffRec \cite{wang2023diffusion}, the model structure is chosen from $\{[ 200 , 600 ], [ 1000 ]\}$ and the step embedding size is 10. The diffusion and inference steps are $\{5, 40, 100\}$ and 0, respectively. The noise scale, minimum and maximum values are chosen from $[1\text{e}^{-5}, 1\text{e}^{-4}, 1\text{e}^{-3}, 1\text{e}^{-2}]$, $[5\text{e}^{-4}, 1\text{e}^{-3}, 5\text{e}^{-3}]$, and $[5\text{e}^{-3}, 1\text{e}^{-2}]$, respectively. To obtain optimal performance, we set the number of clusters to 1. For CVGA \cite{zhang2023revisiting}, the encoder adopts a single-layer lightweight graph convolution \cite{he2020lightgcn}, and the decoder uses a layer of neural network with a consistent dimension of the embedding size of discriminative methods. The dropout setting is consistent with that used in Mult-VAE.

For \textsf{ProEx}, the number of profiles $K$ is fixed to 4, including one original profile and three newly generated ones (see Fig. \ref{fig_pro} for details). The number of environments $\mathcal E$ ranges from 2 to 8 (see Fig. \ref{fig_env} for details). The shape parameter $\alpha_k$ required for sampling in the linear combination within each environment (as defined in Eq. \ref{Dirichlet}) is selected from [0.1,0.2,0.3], with 0.1 yielding the best performance in most cases. The temperature coefficient $\tau$ in the contrastive regularization loss $\mathcal L_{\text{reg}}$ (Eq. \ref{reg}) is set to 0.2 by default. For the cross-space alignment loss, we use contrastive loss for discriminative methods, with the temperature coefficient set to 0.2 by default. For generative methods, we adopt the distribution matching approach defined in Eq. \ref{KL_align}, where the value of $\beta$ is selected from [0.0, 1.0] with a step size of 0.1. The weight $\lambda_1$ of the alignment loss is selected from the range [0,1], while the weight of the contrastive regularization loss $\lambda_2$ is chosen from [0,2] with a step size of 0.1. Empirically, we find that setting the weight of the variance term $\lambda_3$ in Eq. \ref{loss} to 1.0 yields good performance in most cases.

\end{document}